\documentclass{emulateapj}
\usepackage{epsfig,natbib}
\usepackage{graphicx}
\usepackage{enumerate}
\usepackage{amsmath}
\usepackage{multirow}
\usepackage{bigdelim}
\usepackage{longtable}
\usepackage{threeparttable}
\usepackage{subfigure}
\usepackage{textcomp}
\usepackage{gensymb}
\usepackage{color}
\usepackage{morefloats}
\usepackage{xspace}
\usepackage[T1]{fontenc}
\usepackage{placeins}

\newcommand{\thisstudy}{A}
\newcommand{\becker}{B}
\newcommand{\ojeda}{C}

\newcommand{\telpfs}{PFS, D15}
\newcommand{\telcorhel}{CORALIE, H12}
\newcommand{\telcorpre}{CORALIE, V16}
\newcommand{\telj}{HIRES}

\newcommand{\starteff}{\ensuremath{5475 \pm 60\xspace}}
\newcommand{\starlogg}{\ensuremath{4.27 \pm 0.05\xspace}}
\newcommand{\starfe}{\ensuremath{0.36 \pm 0.05\xspace}}
\newcommand{\starvsini}{\ensuremath{1.80^{+0.24}_{0.16}\xspace}}
\newcommand{\starmass}{\ensuremath{0.99 \pm 0.05\xspace}}
\newcommand{\starrad}{\ensuremath{1.18 \pm 0.08\xspace}}

\newcommand{\perb}{\ensuremath{4.1591287 \pm  0.000019\xspace}}
\newcommand{\tcb}{\ensuremath{2457007.932131 \pm 0.000023\xspace}}
\newcommand{\rprsb}{\ensuremath{0.10186 \pm 0.00023\xspace}}
\newcommand{\perd}{\ensuremath{9.03081 \pm 0.00074\xspace}}
\newcommand{\tcd}{\ensuremath{2457006.36927 \pm 0.00044\xspace}}
\newcommand{\rprsd}{\ensuremath{0.02886 \pm 0.00047\xspace}}
\newcommand{\pere}{\ensuremath{0.789597 \pm 0.000013\xspace}}
\newcommand{\tce}{\ensuremath{2457011.34849 \pm 0.00038\xspace}}
\newcommand{\rprse}{\ensuremath{0.01456 \pm 0.00024\xspace}}
\newcommand{\aub}{\ensuremath{0.05047 \pm 0.00085\xspace}}
\newcommand{\auc}{\ensuremath{1.382 \pm 0.023\xspace}}
\newcommand{\aue}{\ensuremath{0.01667 \pm 0.00028\xspace}}
\newcommand{\aud}{\ensuremath{0.0846 \pm 0.0014\xspace}}

\newcommand{\since}{\ensuremath{4043 \pm 593}\xspace}
\newcommand{\sincb}{\ensuremath{441 \pm 65}\xspace}
\newcommand{\sincd}{\ensuremath{157 \pm 23}\xspace}
\newcommand{\sincc}{\ensuremath{0.59 \pm 0.09}\xspace}

\newcommand{\rpb}{\ensuremath{13.11 \pm 0.89\xspace}}
\newcommand{\rpe}{\ensuremath{1.87 \pm 0.13\xspace}}

\newcommand{\gammacorhel}{\ensuremath{-27070.3 \pm 5.1\xspace}}
\newcommand{\jitj}{\ensuremath{3.7 \pm 0.6\xspace}}
\newcommand{\tcc}{\ensuremath{2455992 \pm 10\xspace}}
\newcommand{\perc}{\ensuremath{595.7 \pm 5.0\xspace}}
\newcommand{\omc}{\ensuremath{136 \pm 12\xspace}}
\newcommand{\omb}{\ensuremath{91^{+183}_{-39}}\xspace}
\newcommand{\jitcorhel}{\ensuremath{5.9 \pm 3.5\xspace}}
\newcommand{\mpc}{\ensuremath{411 \pm 18\xspace}}
\newcommand{\mpe}{\ensuremath{9.11 \pm 1.17\xspace}}
\newcommand{\rhopb}{\ensuremath{0.87 \pm 0.18\xspace}}
\newcommand{\mpb}{\ensuremath{356 \pm 12\xspace}}
\newcommand{\rhope}{\ensuremath{7.63 \pm 1.90\xspace}}
\newcommand{\rhopd}{\ensuremath{1.36 \pm 0.42\xspace}}

\newcommand{\mpd}{\ensuremath{12.75 \pm 2.70\xspace}}
\newcommand{\gammapfs}{\ensuremath{20.5 \pm 2.9\xspace}}
\newcommand{\gammaj}{\ensuremath{6.4 \pm 1.5\xspace}}
\newcommand{\eb}{\ensuremath{0.0036^{+0.0049}_{-0.0026}}\xspace}
\newcommand{\gammacorpre}{\ensuremath{-27085.3 \pm 2.7\xspace}}
\newcommand{\ke}{\ensuremath{6.34 \pm 0.78\xspace}}
\newcommand{\kc}{\ensuremath{32.62 \pm 1.14\xspace}}
\newcommand{\kb}{\ensuremath{142.34 \pm 0.85\xspace}}
\newcommand{\jitcorpre}{\ensuremath{6.7 \pm 3.3\xspace}}
\newcommand{\kd}{\ensuremath{3.94 \pm 0.82\xspace}}
\newcommand{\jitpfs}{\ensuremath{6.3 \pm 1.2\xspace}}
\newcommand{\ec}{\ensuremath{0.27 \pm 0.04\xspace}}
\newcommand{\rpd}{\ensuremath{3.71 \pm 0.26\xspace}}

\makeatletter

\newcommand{\Rmnum}[1]{\expandafter\@slowromancap\romannumeral #1@}
\newcommand{\mearthsini}{M$_{\oplus}$sin$i$\xspace}

\newcommand{\msun}{$M_{\odot}$\xspace}
\newcommand{\rsun}{$R_{\odot}$\xspace}
\newcommand{\mearth}{$M_{\oplus}$\xspace}
\newcommand{\rearth}{$R_{\oplus}$\xspace}
\newcommand{\kms}{km~s$^{-1}$\xspace}

\newcommand{\mytilde}{\raise.17ex\hbox{$\scriptstyle\mathtt{\sim}$}}

\newcommand{\shk}{$S_\mathrm{HK}$\xspace}
\newcommand{\wasp}{WASP-47\xspace}
\newcommand{\ms}{m\,s$^{-1}$\xspace}

\newcommand{\Mstar}{\ensuremath{M_{\star}}\xspace}
\newcommand{\Rstar}{\ensuremath{R_{\star}}\xspace} 
 
\newcommand{\fe}{$[$Fe/H$]$\xspace}
\newcommand{\Teff}{$T_{\mathrm{eff}}$\xspace}  
\newcommand{\logg}{\ensuremath{\log g}\xspace} 
\newcommand{\vsini}{\ensuremath{v \sin i}\xspace} 
\newcommand{\gcc}{g\,cm$^{-3}$\xspace}

\newcommand{\Mp}{\ensuremath{M_p}\xspace} 
\newcommand{\Rp}{\ensuremath{R_p}\xspace}

\newcommand{\rhop}{\ensuremath{\rho_p}\xspace}
 
\newcommand{\Kepler}{\textit{Kepler}\xspace} 
\newcommand{\ktwo}{\textit{K2}\xspace}
\newcommand{\mJup}{$M_\mathrm{Jup}$\xspace}
\newcommand{\sinc}{$S_{\mathrm{inc}}$\xspace}
\newcommand{\searth}{$S_{\oplus}$\xspace}

\newcommand{\rprs}{\ensuremath{R_p/R_\star}\xspace}

\makeatother

\shortauthors{Sinukoff}
\shorttitle{WASP-47}

\begin{document}

\pagenumbering{arabic}

\title{Mass constraints of the WASP-47 planetary system from Radial Velocities}

\author{
Evan Sinukoff\altaffilmark{1,14}, 
Andrew W.\ Howard\altaffilmark{2},
Erik A.\ Petigura\altaffilmark{3,15},
Benjamin J.\ Fulton\altaffilmark{1,16},
Howard Isaacson\altaffilmark{4},
Lauren M.\ Weiss\altaffilmark{5},
John M.\ Brewer\altaffilmark{6},
Brad M.\ S.\ Hansen\altaffilmark{7},
Lea Hirsch\altaffilmark{4}, 
Jessie L. Christiansen\altaffilmark{8},
Justin R.\ Crepp\altaffilmark{9},
Ian J.\ M.\ Crossfield\altaffilmark{10,17},
Joshua E.\ Schlieder\altaffilmark{8,18},
David R.\ Ciardi\altaffilmark{8},
Charles A.\ Beichman\altaffilmark{8},
Heather A.\ Knutson\altaffilmark{3},
Bjoern Benneke\altaffilmark{3}, 
Courtney D.\ Dressing\altaffilmark{3,17},
John H.\ Livingston\altaffilmark{11},
Katherine M.\ Deck\altaffilmark{3},
S\'ebastien L\'epine\altaffilmark{12}
Leslie A.\ Rogers\altaffilmark{13}
}




\altaffiltext{1}{Institute for Astronomy, University of Hawai`i at M\={a}noa, Honolulu, HI 96822, USA} 
\altaffiltext{2}{Cahill Center for Astrophysics, California Institute of Technology, 1216 East California Boulevard, Pasadena, CA 91125, USA}
\altaffiltext{3}{Division of Geological and Planetary Sciences, California Institute of Technology, 1255 East California Blvd, Pasadena, CA 91125, USA}
\altaffiltext{4}{Astronomy Department, University of California, Berkeley, CA, USA}
\altaffiltext{5}{Institut de Recherche sur les Exoplan\`etes, D\`epartement de Physique, Universit\`e de Montr\`eal, C.P.\ 6128, Succ.\ Centre-ville, Montr\'eal, QC H3C 3J7, Canada}
\altaffiltext{6}{Department of Astronomy, Yale University and 260 Whitney Avenue, New Haven, CT 06511, USA}
\altaffiltext{7}{Department of Physics \& Astronomy and Institute of Geophysics \& Planetary Physics, University of California Los Angeles, Los Angeles, CA 90095, USA}
\altaffiltext{8}{NASA Exoplanet Science Institute, California Institute of Technology, 770 S. Wilson Ave., Pasadena, CA, USA}
\altaffiltext{9}{Department of Physics, University of Notre Dame, 225 Nieuwland Science Hall, Notre Dame, IN, USA}
\altaffiltext{10}{Department of Astronomy \& Astrophysics, University of California Santa Cruz, 1156 High St., Santa Cruz, CA, USA}
\altaffiltext{11}{Department of Astronomy, The University of Tokyo, 7-3-1 Bunkyo-ku, Tokyo 113-0033, Japan}
\altaffiltext{12}{Department of Physics and Astronomy, Georgia State University, GA, USA}
\altaffiltext{13}{Department of Astronomy \& Astrophysics, University of Chicago, 5640 South Ellis Avenue, Chicago, IL 60637, USA}


\altaffiltext{14}{NSERC Postgraduate Research Fellow}
\altaffiltext{15}{Hubble Fellow}
\altaffiltext{16}{NSF Graduate Research Fellow}
\altaffiltext{17}{NASA Sagan Fellow}
\altaffiltext{18}{NASA Postdoctoral Program Fellow}

\begin{abstract}
    We report precise radial velocity (RV) measurements of WASP-47, a G star that hosts three transiting planets in close proximity (a hot Jupiter, a super-Earth and a Neptune-sized planet) and a non-transiting planet at 1.4 AU.  Through a joint analysis of previously published RVs and our own Keck-HIRES RVs, we significantly improve the planet mass and bulk density measurements. For the super-Earth WASP-47e ($P$ = 0.79 days), we measure a mass of \mpe\,\mearth, and a bulk density of \rhope\,\gcc, consistent with a rocky composition. For the hot Jupiter WASP-47b ($P$ = 4.2 days), we measure a mass of \mpb\,\mearth (1.12 $\pm$  0.04 \mJup) and constrain its eccentricity to $<0.021$ at 3-$\sigma$ confidence. For the Neptune-size planet WASP-47d ($P$ = 9.0 days), we measure a mass of \mpd\,\mearth, and a bulk density of \rhopd\,\gcc, suggesting it has a thick H/He envelope. For the outer non-transiting planet, we measure a minimum mass of \mpc\,\mearth (1.29 $\pm$ 0.06 \mJup), an orbital period of \perc~days, and an orbital eccentricity of \ec. Our new measurements are consistent with but 2--4$\times$ more precise than previous mass measurements.
\end{abstract}
    
\section{Introduction}

Approximately 1\% of Sun-like stars host giant planets on short-period orbits (P $<$ 10 days), known as hot Jupiters \citep[HJs,][]{Howard12, Wright12a}. These planets are thought to have migrated to their observed locations from beyond the ice-line at several AU. One proposed migration mechanism involves dynamical interaction between the planet and protoplanetary disk \citep[e.g.][]{Lin96}. In this case, the planet maintains a low eccentricity. Other ``high-eccentricity migration'' (HEM) modes have been proposed including planet-planet scattering \citep[e.g.][]{Rasio96}, Kozai oscillations induced by either a nearby star \citep[e.g.][]{Wu03} or planet \citep[e.g.][]{Naoz11}, and secular interactions \citep[e.g.][]{Wu11}. In the HEM scenario, gravitational perturbations excite planets onto eccentric orbits, which subsequently shrink and circularize due to stellar tides. Other proposed dynamical effects include misalignment between the orbital axis of the HJ and the stellar spin axis, as well as the destabilization of close-in planets encountered upon migration. 

Observations of systems with HJs are difficult to reconcile with HEM theory. For example, \citet{Schlaufman16} found that HJ host stars are no more likely to host additional giant planets than stars with giant planets at $P > 10$ days. \citet{Knutson14} found no difference between the occurrence of additional giant planets at 1--20 AU in systems with HJs whose orbits are eccentric or misaligned versus circular and aligned with the stellar spin. Moreover, \citet{Dawson15} concluded that the number of migrating Jupiters on highly eccentric orbits is lower than predicted by HEM theory \citep{Socrates12}.  

In support of HEM theory, \citet{Steffen12} found an absence of HJs in close proximity to smaller planets (0.7--5 \rearth) discovered by $Kepler$. However, it remains unclear whether HJs are intrinsically lonely or if their close neighbors have merely evaded detection. For example, \citet{Batygin16} proposed a mechanism for in-situ formation of HJs, which predicts a population of small planets mutually inclined to the HJ, and therefore unlikely to transit.  While HJs are observed to be lonely,  \citet{Huang16} found that roughly half of transiting ``warm-Jupiters" ($P$ = 10--200 days) are accompanied by transiting planets $\sim$ 2--6\,\rearth on interior orbits $P$ $<$ 50 days.  They proposed that the warm-Jupiters in these multi-planet systems formed in-situ and that occasionally this same mechanism might produce a very small fraction of HJs.  These latest theories add to the diversity of theories to explain HJ formation.

WASP-47 is the first star known to host a Jovian-size planet with $P$ $<$ 10 days and additional close-in planets---proof that not all HJs are isolated and strengthening the argument that HEM alone cannot produce the entire population of HJs. The Jovian-size planet WASP-47b orbits the star every 4.2 days. It was first reported and confirmed by \citet{Hellier12} who detected both its transit and radial velocity (RV) signatures. \citet{Becker15} detected two additional transiting planets using \ktwo photometry. One of these planets, WASP-47e, is an ultra-short-period (USP) super-Earth ($P$ = 0.79 days). WASP-47d is Neptune-size ($P$ = 9.0 days). \citet{Becker15} detected transit timing variations (TTVs) of both planets. Their TTV signals are anticorrelated and have a super-period consistent with 52.67-days --- the expected super-period for two such planets near 2:1 orbital mean-motion-resonance \citep{Lithwick12b}.  \citet{Becker15} reported planet mass constraints $M_b$ = $341^{+73}_{-55}$\,\mearth, $M_d$ = $15.2\pm{7}$\,\mearth, and $M_e$ $<$ $22$\,\mearth based on dynamical fits to the observed transit times.  Measurements of the Rossiter-McLaughlin effect by \citep{Ojeda15} ruled out orbits that are strongly misaligned with the stellar spin axis. \citet{Crossfield16} independently validated the planetary system by demonstrating that the star is unlikely to be a blend of multiple stars, via Keck-NIRC2 adaptive optics images and a search for secondary lines in the stellar spectrum.       

A fourth planet, WASP-47c, was detected with an orbital period of $572\pm7$ days by \citet{VanMalle16} from 32 RV observations with the Euler/CORALIE instrument spanning almost 3 years\footnote{WASP-47d and WASP-47e were published before WASP-47c, which was named while the work of \citet{VanMalle16} was still under revision.}. \citet{VanMalle16} measure a minimum mass $\ensuremath{M_c\sin{i}}\xspace = 394 \pm 70$\,\mearth. WASP-47c joins the population of giant planets beyond 1 AU that have been found in systems with HJs \citep{Knutson14}.

WASP-47d and WASP-47e are examples of super-Earth- and Neptune-size planets, which are common around Sun-like stars \citep{Howard12, Fressin13, Petigura13b, Burke15}. Only a handful of these planets have precisely measured masses and bulk densities. Compositional trends have emerged from this limited sample. Planets smaller than $\approx$ 1.6\,\rearth typically have high densities consistent with Earth-like bulk compositions, while most larger planets have low densities that require thick envelopes of H/He \citep{Weiss14, Marcy14a, Lopez14, Rogers15, Dressing15a}. However, there is significant scatter about the mean mass-radius relationship, indicating compositional diversity, even for a fixed planet radius. Due to the limited number of known sub-Neptunes with bright host stars, mass measurements are scarce, and this compositional diversity has yet to be fully explored. 

\citet{Dai15} obtained 26 RVs of \wasp with the Carnegie Planet Finder Spectrograph (PFS), measuring $M_b$ = $370\pm{29}$\,\mearth, $M_e$ = $12.2$ $\pm$ $3.7$\,\mearth, and $M_d$ = $10.4$ $\pm$ $8.4$\,\mearth, consistent with TTV measurements by \citet{Becker15}. \citet{Dai15} measure a bulk density of WASP-47e of 11.2 $\pm$ 3.6\,\gcc, consistent with a rocky and potentially iron-rich composition. Their $\sim$ 80\% measurement uncertainty on the mass of planet d is insufficient to constrain the planet's bulk composition.  

Here we present improved mass constraints of all four planets in the \wasp system by combining Keck-HIRES RVs with the previously published RVs of \citet{Hellier12}, \citet{Dai15}, and \citet{VanMalle16}. This work is part of a NASA ``Key Project'' to measure \ktwo planet masses using Keck-HIRES. Section \ref{sec:obs} of this manuscript summarizes our Doppler observations and spectroscopic constraints of stellar parameters. Our analysis of the RV time-series and resulting planet mass measurements are detailed in Section \ref{sec:analysis}.  In Section \ref{sec:discussion}, we discuss possible compositions of WASP-47e and WASP-47d, eccentricity constraints of the Hot-Jupiter, and interpret these in the context of planet formation and evolution.



\section{Observations}
\label{sec:obs}

\subsection{\ktwo Photometry}
WASP-47 was observed by the Kepler Telescope for 69 consecutive days in Campaign 3 (C3) of NASA's \ktwo mission \citep{Howell14}. It was one of only 55 targets in \ktwo Campaign 3 that was observed in short-cadence mode (60 sec), enabling precise measurement of transit parameters. We adopt the orbital ephemerides, and transit depths reported by \citet{Becker15}.

\subsection{Radial Velocity Measurements}
We collected RV measurements of \wasp using HIRES \citep{Vogt94} at the W.\ M.\ Keck Observatory from 2015 August 10 UT to 2016 October 7 UT (424 days). We followed standard procedures of the California Planet Search \citep[CPS;][]{Howard10}. For each RV observation, we used the  ``C2'' decker ($0\farcs87$ $\times$ 14\arcsec{} slit), which yields a spectral resolution $R$ = 55,000 and is long enough for sky subtraction. Before the starlight entered the spectrometer slit, it first passed through a cell of iodine gas, which imprints a dense set of molecular absorption lines on the stellar spectrum. These iodine lines were used for wavelength calibration and PSF reference. We used an exposure meter to terminate exposures after reaching a SNR per pixel of $\sim$100 at 550\,nm (typically $\sim$15 min). A single iodine-free spectrum was obtained as a stellar template using the ``B3'' decker ($0\farcs57$ $\times$ 14\arcsec{} slit). RVs were measured by forward modeling each observed spectrum as the product of an RV-shifted iodine-free spectrum and a high-resolution/high-SNR iodine transmission spectrum.  The latter was first convolved with an instrumental PSF, modeled as the sum of 13 Gaussians with fixed centers and widths but variable amplitudes \citep{Marcy92, Valenti95, Butler96, Howard09}. Our measured RVs are listed in Table \ref{tb:RVdata}. 

\subsection{Stellar Parameters}
We measured the effective temperature (\Teff), surface gravity (\logg), and metallicity (\fe) of WASP-47 from our iodine-free HIRES spectrum using the updated SME analysis of \citet{Brewer16}. This new methodology yields \logg\, values that are accurate to 0.05~dex, as determined from careful comparisons against stars with \logg determined from asteroseismology \citep{Brewer15}. We find \Teff $= 5475 \pm 60$~K, $\logg = 4.27 \pm 0.05$~dex, and \fe~$= 0.36 \pm 0.05$~dex. To estimate the stellar mass and radius, we fit our spectroscopic measurements of \Teff, \logg, \& \fe to a grid of models from the Dartmouth Stellar Evolution Database \citep{Dotter08} using the \texttt{isochrones} Python package \citep{Morton15a} with uncertainties determined by the \texttt{emcee} Markov Chain Monte Carlo (MCMC) package \citep{Foreman-Mackey13}.  The derived stellar mass and radius are 0.99 $\pm$ 0.03\,\msun and \starrad\,\rsun.  These are consistent with the measurements of 1.04 $\pm$ 0.08\,\msun and 1.15 $\pm$ 0.04\,\rearth by \citet{Mortier13}.  Following \citet{Sinukoff16}, we conservatively adopt uncertainties of 5\% on stellar mass to account for the intrinsic uncertainties of the Dartmouth models estimated by \citet{Feiden12}. 


Following the prescription of \citet{Isaacson10}, we measure \shk indices from the HIRES spectra, which serve as a proxy for stellar activity. Our \shk measurements are listed in Table \ref{tb:RVdata}. The median \shk index of 0.132 is consistent with other inactive stars in the California Planet Search \citep{Isaacson10}. Consistent with this picture, we measure the stellar jitter to be \jitj\,\ms (Table \ref{tb:params}).

\section{Analysis}

\label{sec:analysis}

\begin{deluxetable}{lrrrr}
\tablecaption{RV datasets}
\tablehead{\colhead{Reference\tablenotemark{a}} & \colhead{Instrument} & \colhead{$N_\mathrm{RV}$} &  \colhead{Median Unc.} & \colhead{$\Delta t$}\\
\colhead{} & \colhead{} & \colhead{} & \colhead{[\ms]} & \colhead{[days]}} \\
\startdata
This study & HIRES & 47\tablenotemark{b} & 1.8 & 424 \\
V16 & CORALIE & 26\phantom{a} & 11.4 & 745 \\
D15 & PFS & 26\phantom{a} & 3.1 & 12 \\
H12 & CORALIE & 19\phantom{a} & 11.0 & 560
\enddata
\tablenotetext{a}{V16: \citet{VanMalle16}, D15: \citet{Dai15}, H12: \citet{Hellier12}}
\tablenotetext{b}{We made 74 RV measurements with Keck-HIRES, but omit 17 RVs measured during the WASP-47b transit event on 2015 August 10 UT. We binned the remaining 12 RVs from that night into two measurements for a total of 47 RVs}
\label{tb:RVdatasets}
\end{deluxetable}

\subsection{Radial Velocity Data Analysis}
\label{sec:RVanalysis}

We analyzed the RV time-series using the RV fitting package \texttt{RadVel} (Fulton \& Petigura, in prep.), which is publicly-available on GitHub\footnote{https://github.com/California-Planet-Search/radvel \\ http://radvel.readthedocs.io/en/master/index.html}. We fit our Keck-HIRES RVs along with previously published RV datasets \citep{Hellier12,Dai15,VanMalle16}, summarized in Table \ref{tb:RVdatasets}. We omit the six RV measurements reported by \citet{VanMalle16} that were taken after a CORALIE instrument upgrade.  These would have added two free parameters to our RV model, which was not worth the negligible gain in RV measurements.  After omitting the 17 HIRES observations JD = 2457244.9366--2457245.07451, taken during a WASP-47b transit, we still have 12 out-of-transit observations from that night.  RVs have astrophysical and instrumental errors that manifest on a variety of timescales from minutes to year. Therefore, the consecutive measurements during the same night don't constitute independent measurements. To guard against these data from having a disproportionate influence influence on the fit, we bin the 8 pre-transit RV measurements and bin the 4 post-transit measurements. We note that an analysis of our HIRES RVs alone gives the same planet masses to within 1$\sigma$.

We adopt a four-planet model that is the sum of four Keplerian components. For each of the four datasets, our model includes an RV offset, $\gamma$, as well as an RV ``jitter'' parameter, $\sigma_{\mathrm{jit}}$, to account for additional Doppler noise of astrophysical or instrumental origins. 

Our likelihood function for this analysis follows that of \citet{Howard14}:

\begin{align}
    \begin{split}
        \ln{\mathcal{L}} ={} -\sum_{i} & \left[\frac{\left(v_i-v_m(t_i)\right)^2}{2\left(\sigma_i^2+\sigma_{\mathrm{jit}}^2\right)}\right. \\
        {} & + \ln{\sqrt{2\pi\left(\sigma_i^2 + \sigma_{\mathrm{jit}}^2\right)}}\left.\vphantom{\frac{\left(v\right)^2}{\left(v\right)}}\right],
    \end{split} 
\end{align}
where $v_i$ and $\sigma_i$ are the \textit{i}th RV measurement and corresponding uncertainty, and $v_m$($t_i$) is the Keplerian model velocity at time $t_i$.  To increase the rate of convergence and to to counter the bias toward non-zero eccentricity \citep{Lucy71}, we adopt the following parametrization of our model RV curve: \{$P, T_c, \sqrt{e}\cos{\omega}, \sqrt{e}\sin{\omega}, K$\}, where $P$ is orbital period, $T_c$ is the time of conjunction, $e$ is the orbital eccentricity, $\omega$ is the longitude of periastron and $K$ is the RV semi-amplitude. 

We first find the maximum-likelihood model using the minimization technique of \citet{Powell64}, then perturb the best-fitting free parameters by up to 3\% to start 100 parallel MCMC chains. The free parameters of the RV model are adopted as the MCMC step parameters.  \texttt{RadVel} incorporates the affine-invariant sampler of the \texttt{emcee} package \citep{Foreman-Mackey13}. The Gelman-Rubin \citep{Gelman92} and $T_{z}$ statistics \citep{Ford06} are checked in real-time during the MCMC exploration and the chains are deemed well-mixed and the MCMC run is halted when the Gelman-Rubin is within 3\% of unity and $T_{z}$ $>$ $1000$ for all free parameters.  

\begin{figure*}
    \centering
    \includegraphics[width=1.5\columnwidth]{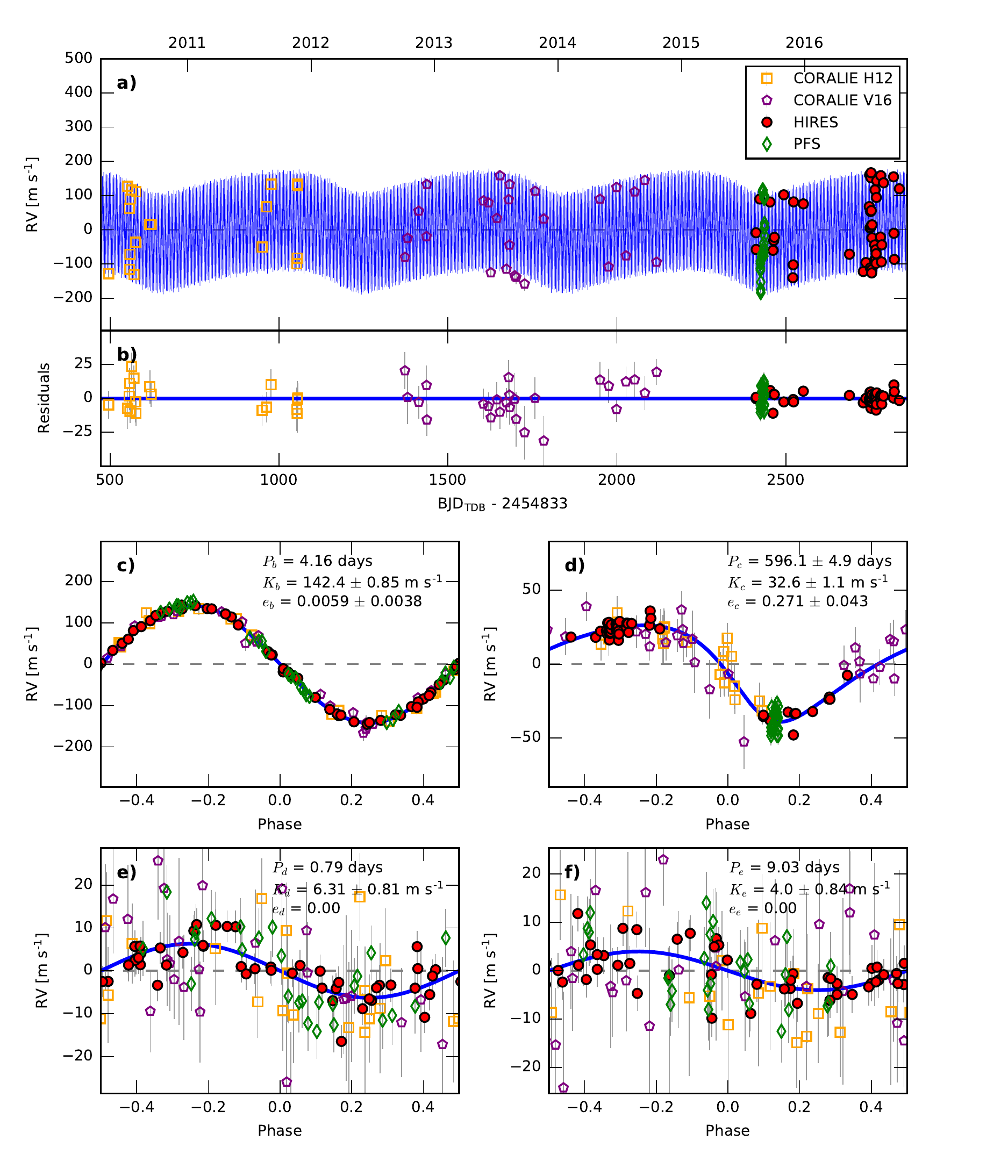}
    \caption{Four-planet RV model of WASP-47, assuming circular orbits for WASP-47d and WASP-47e \textbf{a)} The RV time-series. Filled red circles indicate Keck-HIRES data. Orange squares represent CORALIE data published by \citet{Hellier12}. Purple pentagons represent CORALIE data published by \citet{VanMalle16}. Green diamonds indicate PFS data published by \citet{Dai15}. The solid blue line corresponds to the most likely model. Note that the orbital parameters listed in Table \ref{tb:params} are the median values of the posterior distributions. Error bars for each independent dataset include an RV jitter term listed in Table \ref{tb:params}, which are added in quadrature to the measurement uncertainties. \textbf{b)} Residuals to the maximum-likelihood fit. \textbf{c-f)} The RV time-series phase folded at the orbital periods of each of the four planets after subtracting the other three planet signals.}
    \label{fig:RVfit}
\end{figure*}

We assume circular orbits for WASP-47d and WASP-47e while allowing the eccentricities of WASP-47b and WASP-47c to vary freely. An N-body dynamical stability analysis by \citet{Becker15} showed that the orbits of the inner three planets are unstable when eccentricities of the three inner planets exceed $\sim$ $0.05$. For the $\sim$ 4--6\,\ms RV signals of WASP-47d and WASP-47e, our signal-to-noise is too low to distinguish between eccentricities of 0.00 and 0.05. The orbital periods and orbital phases of WASP-47b, d, and e were locked at the values reported in \citet{Becker15} from transits. We adopt uninformed priors (i.e. no priors) on all free step parameters and step in linear parameter space. The median values and the 68\% credible intervals are reported in Table \ref{tb:params}. The best-fitting RV model is shown in Figure \ref{fig:RVfit} 

We searched for additional companions at large orbital distances by testing RV models with and without a constant radial acceleration term, $\mathrm{d}v/\mathrm{d}t$. We compared these two models using the Bayesian Information Criterion (BIC), with the RV jitter fixed at the values in Table \ref{tb:params}. We compute $\Delta${}BIC = BIC$_{\mathrm{d}v/\mathrm{d}t}-$BIC$_{\mathrm{d}v/\mathrm{d}t=0}$ = 3.8, indicating that the simpler model is preferred, so we adopt $\mathrm{d}v/\mathrm{d}t$ = 0.  

We investigated whether the Keplerian orbit approximation is valid for our RV model, given the dynamical influences of the three inner planets on each other.  First, we considered the TTV amplitudes, which indicate the order of magnitude of non-Keplerian effects.  The TTV amplitudes of planets b, d, and e measured by \citet{Becker15} of 0.63 min, 7.3 min, and < 1.2 min are 0.01\%, 0.06\%, and $<$ 0.1\% of the respective orbital periods.  We assessed whether these deviations from Keplerian orbits are significant given the precision of our RV measurements.  Given RV semi-amplitude $K$ and assuming a phase shift equal to the TTV amplitude $\Delta{T}$, the deviation of RV($t$) is: 
\begin{equation} \label{eq1}
\begin{split}
\Delta\text{RV(\text{$t$})} & = \frac{\partial\text{RV}}{\partial t}\Delta{T} \\
 & = \frac{2\pi\text{$K$}}{P}\cos\left(\frac{2\pi\text{$t$}}{P}\right)\Delta{T}.
\end{split}
\end{equation}
The maximum $\Delta\text{RV}$ is $2\pi{K}P^{-1}\Delta{T}$, which evaluates to 0.09\,\ms, 0.01\,\ms, and $<$ 0.03\,\ms for planets  b, d, and e respectively.  These represent upper bounds to the orbit-averaged deviations from Keplerian over the K2 time baseline.  These deviations are much smaller than our RV measurement uncertainties (1.5--2.0 m/s).

Since the RV time-series is much longer than the K2 baseline, one may wonder if there are large amplitude deviations from Keplerian orbits that build up over longer timescales.  To verify that the TTVs remain small over the timescale of RV observations, we used the symplectic N-body integrator TTVFast \citep{Deck14} to numerically integrate the planet orbits over 2000 days.  The orbital elements were initialized at the maximum-likelihood solution obtained from RVs.  The TTV amplitudes of planets b, d, and e remained at 0.6 min, 7 min and $<$ 1 min respectively over the 2000 day timespan.

We note that the orbital periods of planets b and d measured by \citet{Becker15} do not accurately reflect the average orbital periods that would be measured over many years.  \citet{Becker15} measured $P$ by fitting a linear ephemeris to the \ktwo transits.  Since the \ktwo photometry only spans one TTV super-period, the \citet{Becker15} orbital periods could be different from the average orbital periods over the time baseline of our RV measurements, which spans many TTV super-periods. 

To quantify the additional uncertainties of average orbital periods, we used the 2000-day baseline of transit times obtained with TTVFast.  For each planet, we performed a linear fit to every unique set of $N$ consecutive transit times, where $N$ is the number of transits observed in the K2 photometry.  The resulting distribution of slopes (orbital periods) provides an estimate of the uncertainty of the average orbital period attributed to the limited timescale of \ktwo observations. The 1-sigma uncertainties obtained from these orbital period distributions are $\pm$ 0.000019 days and $\pm$ 0.00074 days for planets b and d respectively.  These are $\sim$ 4$\times$ larger than the uncertainties reported by \citet{Becker15}.  We refit our RV time-series using these larger orbital period uncertainties, but there was no change in the RV solution or corresponding uncertainties. The scale of these uncertainties is still a tiny fraction of the RV phase.  Nevertheless we recommend that future studies adopt these larger uncertainties on average orbital period, which are listed in Table \ref{tb:params}.

\begin{deluxetable}{lrlr}
\tablecaption{WASP-47 system parameters}
\tablehead{\colhead{Parameter} & \colhead{Value} & \colhead{Units} & \colhead{Ref.}}
\startdata
\hline
\sidehead{{\bf Stellar Parameters}} 
\Teff  &   \starteff    & $K$     & \thisstudy \\
\logg  &   \starlogg    & dex     & \thisstudy \\
\fe    &   \starfe      & dex     & \thisstudy \\ 
\vsini &   \starvsini   & \kms    & \ojeda     \\
\Mstar &   \starmass    & \msun   & \thisstudy \\
\Rstar &   \starrad     & \rsun   & \thisstudy \\
\hline 
\sidehead{{\bf Planet Parameters}}
\sidehead{WASP-47b} 
$P$          & \perb   & days        & \thisstudy, \becker       \\
$T_{\rm{conj}}$  & \tcb    & BJD         & \becker       \\
\rprs        & \rprsb  & ---         & \becker       \\
$a$ & \aub & AU & \thisstudy \\
$S_{\rm{inc}}$ & \sincb  & $S_\oplus$  & \thisstudy    \\
\Rp          & \rpb    & \rearth     & \thisstudy    \\
$e$          & \eb     & ---         & \thisstudy    \\
$\omega$     & \omb    & deg         & \thisstudy    \\
$K$          & \kb     & \ms         & \thisstudy    \\
\Mp          & \mpb    & \mearth     & \thisstudy    \\
\rhop        & \rhopb  & \gcc        & \thisstudy    \\
\sidehead{WASP-47c} 
$P$          & \perc   & days        & \thisstudy    \\
$T_{\rm{conj}}$  & \tcc    & BJD         & \thisstudy    \\
$a$ & \auc & AU & \thisstudy \\
$S_{\rm{inc}}$     & \sincc  & $S_\oplus$  & \thisstudy    \\
$e$          & \ec     & ---         & \thisstudy    \\
$\omega$     & \omc    & deg         & \thisstudy    \\
$K$          & \kc     & \ms         & \thisstudy    \\
\Mp          & \mpc    & \mearthsini & \thisstudy    \\
\sidehead{WASP-47d (circular orbit assumed)} 
$P$          & \perd   & days        & \thisstudy,  \becker       \\
$T_{\rm{conj}}$  & \tcd    & BJD         & \becker       \\
\rprs        & \rprsd  & ---         & \becker       \\
$a$ & \aud & AU & \thisstudy \\
$S_{\rm{inc}}$      & \sincd  & $S_\oplus$  & \thisstudy    \\
\Rp          & \rpd    & \rearth     & \thisstudy    \\
$K$          & \kd     & \ms         & \thisstudy    \\
\Mp          & \mpd    & \mearth     & \thisstudy    \\
\rhop        & \rhopd  & \gcc        & \thisstudy    \\
\sidehead{WASP-47e (circular orbit assumed)} 
$P$          & \pere   & days        & \becker       \\
$T_{\rm{conj}}$ & \tce    & BJD         & \becker       \\
\rprs        & \rprse  & ---         & \becker       \\
$a$ & \aue & AU & \thisstudy \\
$S_{\rm{inc}}$      & \since  & $S_\oplus$  & \thisstudy    \\
\Rp          & \rpe    & \rearth     & \thisstudy    \\
$K$          & \ke     & \ms         & \thisstudy    \\
\Mp          & \mpe    & \mearth     & \thisstudy    \\
\rhop        & \rhope  & \gcc        & \thisstudy    \\
\hline
\sidehead{\bf{Other}}
$\gamma_{\rm \telj}$             & \gammaj      & \ms  &  \thisstudy \\
$\gamma_{\rm \telpfs}$           & \gammapfs    & \ms  &  \thisstudy \\
$\gamma_{\rm \telcorhel}$        & \gammacorhel & \ms  &  \thisstudy \\
$\gamma_{\rm \telcorpre}$        & \gammacorpre & \ms  &  \thisstudy \\
$\sigma_{\rm jit, \telj}$        & \jitj        & \ms  &  \thisstudy \\
$\sigma_{\rm jit, \telpfs}$      & \jitpfs      & \ms  &  \thisstudy \\
$\sigma_{\rm jit, \telcorhel}$   & \jitcorhel   & \ms  &  \thisstudy \\
$\sigma_{\rm jit, \telcorpre}$   & \jitcorpre   & \ms  &  \thisstudy \\
\enddata
\tablecomments{$S_{\rm{inc}}$ = Incident flux, $T_{\rm{conj}}$ = Time of conjunction, \thisstudy: This study, \becker: \citet{Becker15}, \ojeda: \citet{Ojeda15}.  H12: \citet{Hellier12}, D15: \citet{Dai15}, V16: \citet{VanMalle16}. Orbital periods of planets b and d are those from \citet{Becker15}, but with larger uncertainties (See \S3).}
\label{tb:params}
\end{deluxetable}

\section{Discussion}
\label{sec:discussion}

\begin{figure}
    \centering
    \includegraphics[width=\columnwidth]{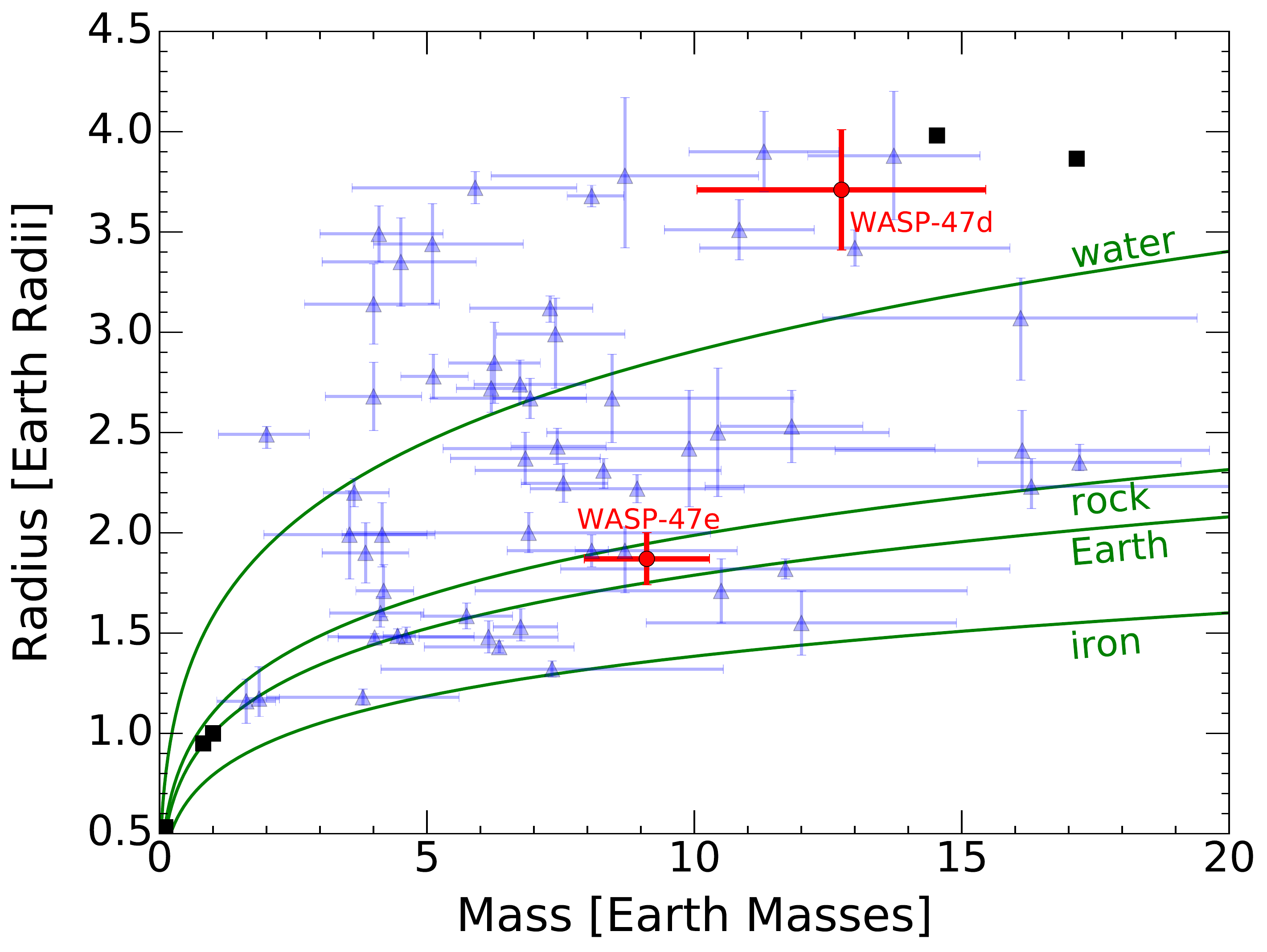}
    \caption{Radii and masses of all confirmed planets whose mass and radius are measured to better than 50\% (2$\sigma$) precision (blue triangles). Solar System planets are represented as black squares. Red circles indicate our measurements of WASP-47d and WASP-47e. Green curves show the expected planet mass-radius curves for 100\% iron, 100\% rock (Mg$_{\rm2}$SiO$_{\rm4}$), 100\% water (ice), and Earth-like (67\% rock, 33\% iron) compositions according to models by \citet{Fortney07}.}
    \label{fig:MR}
\end{figure}

Figure \ref{fig:MR} shows the mass-radius distribution of all confirmed planets with $R_p$ $<$ 4.0\,\rearth whose mass and radius are measured to better than 50\% precision ($2\sigma$) either by RVs or TTVs\footnote{NASA Exoplanet Archive, UT 24 September 2016, http://exoplanetarchive.ipac.caltech.edu}. Previous studies of small planets from the prime \Kepler mission revealed a transition in the typical composition of planets from mostly rocky to planets having thick envelopes of low density H/He at $\approx$ 1.6\,\rearth \citep{Weiss14, Marcy14a, Lopez14, Rogers15, Dressing15a}. An important open question is if and how this transition depends on incident stellar flux. \citet{Hutter16} illustrate that the population of planets $<$ 30 \mearth tend to have fewer volatiles as incident flux increases, consistent with atmospheric loss via photoevaporation. WASP-47e is among the most highly irradiated small planets with a well-measured mass, and thus helps us to probe the mass-radius relationship at extreme values of incident stellar flux, in a regime similar to Kepler-10b, Kepler-78b, and 55 Cnc e. 

The measured mass of \wasp{e} (\mpe\,\mearth) is consistent with the measurement of \citet{Dai15} ($12.2 \pm 3.7$\,\mearth) at the 1$\sigma$ level. We improve the fractional uncertainty from 30\% to 13\%, allowing for a more detailed interpretation of composition.  The measurements of \citet{Dai15} favored an admixture of 50\% iron and 50\% rock.  Assuming an iron-rock admixture, we sample our planet mass and radius posterior distributions and compute an iron mass fraction (IMF) using Equation 8 of \citet{Fortney07}. From 100,000 independent samples, we obtain a median IMF of 13\% and a 1$\sigma$ upper limit of 24\%, suggesting that \wasp{e} is mostly rock.  Its IMF is lower than Earth's IMF (33\%) at 80\% confidence.  Alternatively, \wasp{e} could have an IMF similar to Earth but possess a significant atmosphere of a high mean molecular weight species, such as water or sulfur. 

The measured mass and radius of WASP-47d (\mpd\,\mearth and \rpd\,\rearth) are consistent with several other planets, including Kepler-94b, Kepler-95b, Kepler-30b, KOI-142b, and GJ 3470b. With an incident flux \sinc = \sincd\,\searth, the atmosphere of WASP-47d might have undergone significant photoevaporation. Nevertheless, it must still have an atmosphere containing some amount of H/He.  There are a number of degenerate planet compositions in this region of the mass-radius diagram with different fractions of rock, iron, water, and H/He \citep{Rogers10, Valencia13}. Possible compositions include a small iron-rich or rocky core with an extended H/He or steam envelope, or a water-world with a modest H/He envelope. Future transmission spectroscopy observations would help to break these degeneracies.  



WASP-47e is among the few known USP planets $>$ 1.5\,\rearth.  \citet{Lopez16} explains the dearth of larger USP planets as a consequence of photoevaporation of H/He envelopes of larger planets that formed water-poor.   The one potential counter-example noted by \citet{Lopez16} is the 1.9\,\rearth USP planet 55 Cnc e. The most recent mass and radius constraints suggest the presence of a water-rich envelope, 8 $\pm$ 3\% of the planet's mass.   

55 Cnc has remarkable similarities to WASP-47. It hosts a USP super-Earth (55 Cnc e), a non-transiting giant planet (55 Cnc b) at $P$=15 days, and three additional non-transiting planets at $P$ = 44, 262, and $\sim$ 4800 days.  The fact that these systems host both a HJ and a USP planet suggests that their formations are linked in some way.  Moreover, the mass and radius of 55 Cnc e (8.3 $\pm$ 0.3\,\mearth, 1.92 $\pm$ 0.08\,\rearth) are consistent with WASP-47e (\mpe\,\mearth, \rpe\,\rearth).   Therefore, both planets could have water-rich envelopes.     More well-characterized USP planets $\approx$ 2\,\rearth are needed to determine if they represent a distinct population of USP planets spawning from unique formation and/or evolutionary processes.  In particular, as proposed by \citet{Huang16}, WASP-47b and 55 Cnc b might represent the rare close-in extremes of in-situ formation hypothesized to produce the $\sim$ 50\% of warm-Jupiters ($P$ = 10--200 days) that have small companions at shorter orbital distances.  This highlights the limitations of classifying HJs and warm-Jupiters based on orbital period alone, without taking the more complete system architecture into account.

One clue about the formation history of 55 Cnc e is the fact that it transits whereas the outer planet (P $\sim$4800 days) is claimed, on the basis of HST astrometry, to be inclined to the line of sight by 30 degrees \citep{McArthur04}.  \citet{Hansen15} showed that if 55 Cnc e formed slightly beyond its current orbit, and migrated inwards through tidal dissipation, it would have crossed a pair of secular resonances in the system, which could have boosted its inclination and/or eccentricity.  This would increase the tidal heating and potentially devolatilize the planet or drive it to Roche lobe overflow.  The WASP-47 system also shows a potential secular resonance if WASP-47e once had a semi-major axis of 0.022 AU. Although this system clearly did not experience pumping of the inclination, a small but finite initial eccentricity for WASP-47d could have driven tidal evolution of WASP-47e through this resonance and rapidly increased the tidal heating, potentially leading to strong devolatilization. If WASP-47e and WASP-47d both formed as Neptune-size planets,, but WASP-47e was heated or tidally stripped, then their current difference in densities reflects their evolution rather than their origins.  







From our MCMC analysis of the RV time-series, we determine the orbital eccentricity of the HJ to be $< 0.021$ at $99.7\%$ (3$\sigma$) confidence. The very low eccentricity and the alignment between the orbital axis and stellar spin \citep{Ojeda15} are consistent with disk migration, in-situ formation, and the aforementioned secular interaction scenario.  In future, this eccentricity constraint can be used to inform TTV models. 


WASP-47 has a high metallicity (\starfe\,dex) which has been shown to be associated with HJ occurrence and giant planet occurrence \citep[e.g.][]{Fischer05, Buchhave14}. The Kepler sample of Earth-size planets were found around stars of widely varying metallicity \citep{Buchhave14}. However, if USPs are associated with metal-rich stars, it suggests different formation pathway than the bulk of known Earth-size planets---one that may be more closely associated with HJs.   Although it is beyond the scope of this study, a comparison between the metallicities of stars hosting HJs with those hosting USPs will provide a useful test of the relationship between the formation of USPs and HJs. 

We note that while this manuscript was under review, \citet{Almenara16} reported mass and radius constraints of the WASP-47 system using a photodynamical model.  They simultaneously fit the \ktwo photometry and the RV measurements of \citet{Hellier12}, \citet{Dai15}, and \citet{VanMalle16}.  Their planet mass measurements are consistent with this study at the 1-$\sigma$ level.  Future incorporation of our Keck-HIRES RVs into a photodynamical analysis would further improve constraints of the WASP-47 system.

\begin{deluxetable}{lrrr}[h!]
\tablecaption{Relative radial velocities, Keck-HIRES}
\tablehead{\colhead{BJD} & \colhead{RV} [\ms]\tablenotemark{a} & \colhead{Unc. [\ms]\tablenotemark{b}} & \colhead{\shk}}
\startdata
2457244.871067 & 0.48 & 1.67 & 0.126 \\
2457244.878949 & 5.92 & 1.77 & 0.132 \\
2457244.887016 & -1.84 & 1.66 & 0.133 \\
2457244.895257 & -3.04 & 1.82 & 0.135 \\
2457244.903486 & -1.37 & 1.64 & 0.132 \\
2457244.911878 & -3.26 & 1.84 & 0.127 \\
2457244.920153 & -5.65 & 1.59 & 0.135 \\
2457244.928510 & -8.04 & 1.76 & 0.133 \\
2457244.936600 & -8.56 & 1.66 & 0.135 \\
2457244.944818 & -2.06 & 1.71 & 0.135 \\
2457244.953082 & -0.15 & 1.66 & 0.130 \\
2457244.961427 & -0.13 & 1.62 & 0.132 \\
2457244.969923 & -4.30 & 1.73 & 0.131 \\
2457244.978731 & -9.13 & 1.77 & 0.131 \\
2457244.987215 & -20.02 & 1.73 & 0.130 \\
2457244.995815 & -20.64 & 1.68 & 0.136 \\
2457245.004993 & -29.55 & 1.73 & 0.121 \\
2457245.013720 & -32.44 & 1.66 & 0.127 \\
2457245.022227 & -39.56 & 1.67 & 0.135 \\
2457245.030607 & -39.94 & 1.59 & 0.136 \\
2457245.038917 & -43.69 & 1.62 & 0.104 \\
2457245.047448 & -48.47 & 1.65 & 0.129 \\
2457245.056198 & -53.78 & 1.52 & 0.129 \\
2457245.065700 & -50.95 & 1.79 & 0.114 \\
2457245.074508 & -46.85 & 1.60 & 0.123 \\
2457245.083096 & -44.16 & 1.65 & 0.130 \\
2457245.091812 & -48.39 & 1.52 & 0.133 \\
2457245.100712 & -57.42 & 1.53 & 0.129 \\
2457245.110018 & -52.81 & 1.68 & 0.134 \\
2457256.103458 & 96.66 & 2.11 & 0.135 \\
2457286.030224 & 87.81 & 2.19 & 0.089 \\
2457294.949126 & -53.43 & 2.17 & 0.115 \\
2457296.992830 & -24.59 & 2.00 & 0.137 \\
2457298.980931 & -15.40 & 3.62 & 0.131 \\
2457326.879645 & 108.91 & 1.96 & 0.036 \\
2457353.819776 & -133.61 & 2.05 & 0.142 \\
2457354.803856 & -95.88 & 1.95 & 0.128 \\
2457355.794764 & 88.20 & 1.99 & 0.127 \\
2457384.711392 & 82.45 & 1.93 & 0.127 \\
2457521.108185 & -64.84 & 1.86 & 0.126 \\
2457562.108559 & -115.52 & 1.81 & 0.160 \\
2457570.076758 & -89.15 & 2.05 & 0.151 \\
2457580.060228 & 74.59 & 1.83 & 0.140 \\
2457581.046494 & 167.52 & 1.81 & 0.131 \\
2457582.043488 & 12.38 & 1.98 & 0.135 \\
2457583.061003 & -104.76 & 1.83 & 0.136 \\
2457583.922512 & 13.32 & 1.83 & 0.133 \\
2457584.109907 & 59.97 & 1.80 & 0.127 \\
2457584.914664 & 158.68 & 1.90 & 0.137 \\
2457585.068538 & 173.44 & 1.79 & 0.135 \\
2457585.911306 & 62.57 & 1.98 & 0.140 \\
2457586.089591 & 19.48 & 2.04 & 0.127 \\
2457586.909613 & -110.17 & 1.72 & 0.131 \\
2457587.088454 & -119.61 & 1.67 & 0.138 \\
2457587.950710 & -16.79 & 2.01 & 0.136 \\
2457588.097234 & 21.03 & 1.80 & 0.141 \\
2457595.894592 & -81.36 & 2.24 & 0.138 \\
2457596.120574 & -37.76 & 2.18 & 0.121 \\
2457596.917828 & 123.24 & 2.11 & 0.131 \\
2457598.938091 & -51.78 & 2.07 & 0.119 \\
2457599.106292 & -79.28 & 1.95 & 0.135 \\
2457599.928092 & -91.45 & 1.83 & 0.140 \\
2457600.118064 & -63.52 & 1.87 & 0.135 \\
2457600.927270 & 101.12 & 2.06 & 0.138 \\
2457602.055444 & 149.79 & 2.09 & 0.132 \\
2457612.852525 & -14.58 & 1.83 & 0.136 \\
2457614.023768 & 164.96 & 1.77 & 0.135 \\
2457615.871589 & -87.25 & 1.96 & 0.132 \\
2457616.894897 & -37.49 & 1.90 & 0.133 \\
2457622.042983 & 143.94 & 2.30 & 0.129 \\
2457651.804239 & 161.48 & 1.74 & 0.128 \\
2457652.803402 & -3.55 & 1.77 & 0.145 \\
2457653.938888 & -80.03 & 1.86 & 0.142 \\
2457668.749161 & 126.48 & 2.14 & 0.143
\enddata
\tablenotetext{a}{RVs do not include zero point offset ($\gamma_{\rm \telj}$, Table \ref{tb:params})}
\tablenotetext{b}{Uncertainties do not include jitter ($\sigma_{\rm jit, \telj}$, Table \ref{tb:params})} 
\label{tb:RVdata}
\end{deluxetable}

\acknowledgements
We thank the many observers who contributed to the measurements reported here. We thank Geoff Marcy and Trevor David for helpful discussions. We thank Tom Greene, Michael Werner, Michael Endl, and William Cochrane for participation in our NASA Key Project. We gratefully acknowledge the efforts and dedication of the Keck Observatory staff.  This paper includes data collected by the \ktwo mission. Funding for the \ktwo mission is provided by the NASA Science Mission directorate.  E.~S.\ is supported by a post-graduate scholarship from the Natural Sciences and Engineering Research Council of Canada. E.~A.~P.\ acknowledges support by NASA through a Hubble Fellowship grant awarded by the Space Telescope Science Institute, which is operated by the Association of Universities for Research in Astronomy, Inc., for NASA, under contract NAS 5-26555. B.~J.~F.\ was supported by the National Science Foundation Graduate Research Fellowship under grant No. 2014184874.  A.~W.~H.\ acknowledges support for our \ktwo team through a NASA Astrophysics Data Analysis Program grant.  A.~W.~H.\ and I.~J.~M.~C.\ acknowledge support from the \ktwo Guest Observer Program.  L.~M.~W.\ acknowledges the Trottier Family Foundation for their generous support. This work was performed [in part] under contract with the Jet Propulsion Laboratory (JPL) funded by NASA through the Sagan Fellowship Program executed by the NASA Exoplanet Science Institute. This research has made use of the NASA Exoplanet Archive, which is operated by the California Institute of Technology, under contract with the National Aeronautics and Space Administration under the Exoplanet Exploration Program. Finally, the authors extend special thanks to those of Hawai`ian ancestry on whose sacred mountain of Maunakea we are privileged to be guests.  Without their generous hospitality, the Keck observations presented herein would not have been possible.

{\it Facilities:} \facility{Kepler}, \facility{Keck-HIRES}. 
\FloatBarrier
\bibliographystyle{apj}
\bibliography{references}
\end{document}